# A Benchmark Study on Sentiment Analysis for Software Engineering Research

Nicole Novielli, Daniela Girardi, Filippo Lanubile
University of Bari Aldo Moro, Italy
{nicole.novielli, daniela.girardi, filippo.lanubile}@uniba.it

## ABSTRACT

A recent research trend has emerged to identify developers' emotions, by applying sentiment analysis to the content of communication traces left in collaborative development environments. Trying to overcome the limitations posed by using off-the-shelf sentiment analysis tools, researchers recently started to develop their own tools for the software engineering domain. In this paper, we report a benchmark study to assess the performance and reliability of three sentiment analysis tools specifically customized for software engineering. Furthermore, we offer a reflection on the open challenges, as they emerge from a qualitative analysis of misclassified texts. [1]

## CCS CONCEPTS

• **Collaborative and social Computing** → Collaborative and social computing theory, concepts and paradigms; *Computer supported cooperative work;* • **Information retrieval** → Retrieval tasks and goals; *Sentiment analysis;* • **Software creation and management** → **Collaboration in software development**; *Programming teams*

## 1. INTRODUCTION

Emotion awareness in software engineering is receiving increasing attention as part of human factors of software engineering[2]. As such, a recent research trend has emerged to study developers' emotions as they are shared and conveyed in the communication channels within collaborative development environments, including issue tracking systems (e.g., Jira) [22][27], software repository forges (e.g., GitHub) [15][31][38], and technical Q&A sites (e.g., Stack Overflow)[7].

Sentiment analysis, i.e. the study of the subjectivity (neutral vs. emotionally loaded) and polarity (positive vs. negative) of a text [29], has emerged as the main technique for mining affective states from textual developer-generated content. This choice is also supported by the availability of off-the-shelf sentiment analysis tools, sometimes available as open source software. However, off-the-shelf sentiment analysis tools have been trained on non-technical domains and have been demonstrated to produce unreliable results in software engineering [16]. Trying to overcome the limitations posed by using off-the-shelf sentiment analysis tools, researchers recently started to develop their own tools specifically customized for the software engineering domain [1][3][5][6][18][20][30].

In this study, we perform a replication of the study by Jongeling et al. [16] to assess the performance and reliability of three sentiment analysis tools, which have been specifically optimized for the domain of software development and were not available at the time of the original study. The original study compares the predictions of widely used off-the-shelf sentiment analysis tools, showing not only how these tools disagree with human annotation of developers' communication channels, but also how they disagree with each other. We enhance the state of the art by investigating to what extent fine-tuning sentiment analysis tools for the software engineering (SE) domain do succeed in improving the accuracy of emotion detection.

For this purpose, we reuse and adapt the research questions of the original study to include consideration of SE-specific sentiment analysis tools, as follows:

*RQ1: To what extent do different SE-specific sentiment analysis tools agree with emotions of software developers?*

*RQ2: To what extent do results from different SE-specific sentiment analysis tools agree with each other?*

To address our research questions, we study whether the SE-specific sentiment analysis tools agree with the polarity labels provided by human raters (RQ1) and to what extent they agree with each other (RQ2). Specifically, we assess the performance of three publicly available SE-specific tools, namely Senti4SD [5], SentiStrengthSE [18], and SentiCR [1], and we compare their performance with the baseline represented by Sentistrength [40]. To enable comparison with the original study, we assess the performance on the Jira dataset of comments annotated by Murgia et al. [28] with emotion labels from the Shaver framework [37]. Furthermore, we extend the benchmark by including a gold standard of questions, answers, and comments from Stack Overflow, for which annotation was performed by adopting the same theoretical framework of affect [5].

---

[2] See the SEmotion workshop series on Emotion Awareness in Software Engineering, co-located with ICSE since 2016: http://collab.di.uniba.it/semotion/

Building a reliable gold standard for natural language tasks is a crucial issue and previous research has investigated the reliability of crowdsourced vs. expert annotations [39]. As far as affect labeling is concerned, two labeling approaches can be adopted: model-driven and ad hoc annotation. Model-driven annotation is inspired by theoretical models of affect, which are translated into detailed guidelines and are used as a reference for the human raters, after a preliminary training. It is the case of the Jira and Stack Overflow datasets, mentioned above. Conversely, in ad hoc annotation, the raters are required to provide polarity labels according to their subjective perception of the semantic orientation of the text. A preliminary study by Novielli et al. [26] advocated in favor of using a theoretical model of affect when framing research questions and defining a methodology for empirical studies on emotion awareness in software engineering. However, evidence has not been provided yet about the impact of the labeling approach (model-driven vs. ad hoc annotation) on the quality of gold standards and, thus, on the reliability of sentiment analysis. We then formulate the following additional research question:

*RQ3: To what extent do the labeling approach (model-driven vs. ad-hoc annotation) has an impact on the performance of SE-specific sentiment analysis tools?*

To address RQ3, we compare the performance of SE-specific sentiment analysis tools on the two datasets labeled by leveraging model-driven annotation with performance observed with two additional datasets built by adopting ad hoc annotation [1][20].

As a further contribution, we discuss the open challenges for sentiment analysis in software engineering, as they emerge from a qualitative analysis of the texts that were misclassified by all the three tools in this study.

The remainder of this paper is organized as follows. In Section 2 we report the background on affect modeling and labeling and provide an overview of related work on sentiment analysis in software engineering. In Section 3, we describe the sentiment analysis tools included in the current study. In Section 4, we provide detailed information on the datasets included in the benchmarking and on the experimental setting. Results are provided in Section 5, where we report the performance of sentiment analysis tools and the results of the qualitative analysis of errors. In Section 6, we discuss our findings and the lessons learned from the current study. Finally, we assess the threats to study validity (Section 7) and provide conclusions (Section 8).

## 2. BACKGROUND AND RELATED WORK

### 2.1 Affect Modeling and Labeling

Sentiment analysis is the task of extracting the positive or negative orientation of opinions and emotions conveyed by a text. Mining affective states from text involves, on one hand, to characterize them according to bi-dimensional models. It is the case of the 'circumplex model' of affect, which represents emotions along two axes: (1) the affect polarity, or emotion valence and (2) its level of activation, also known as arousal or intensity [33]. On the other hand, some studies explicitly deal with discrete emotion labeling, by looking for linguistic cues of specific affective states, based on the assumption that a limited set of basic emotions exists, such as joy, happiness, sadness, or anger, although there is no consensus about the nature and the number of these basic emotions [11][19][37].

Regardless of the specific theoretical model, affective states can be easily and intuitively mapped to the polarity dimension, i.e. classified as positive, negative, or neutral. For example, Lazarus [19] describes nine negative (anger, fright, anxiety, guilt, shame, sadness, envy, jealousy, and disgust) and seven positive (happiness, pride, relief, love, hope, compassion, and gratitude) emotions, with their appraisal patterns: positive emotions are triggered if the situation experienced is congruent with an individual goal, otherwise negative emotions are prompted. This holds true also for other states of the affective spectrum, such as opinions or stances, which are traditionally investigated by research in sentiment analysis.

As for affect labeling, at least two different approaches can be adopted for label acquisition, with different degrees of specialization and control. On one hand, we have model-driven annotation, devising an annotation scheme that is grounded on a theoretical framework of affect. *Model-driven annotation* is typically implemented into a rigorous approach involving the use of detailed guidelines and the training of the raters. On the other hand, we have ad hoc annotation, which is based on the raters' subjective perception of the meaning of the affective labels. With *ad hoc annotation*, the raters are typically required to assign a polarity label to each text item, based only on their personal perception of the positive or negative affective load of the text.

### 2.2 Sentiment Analysis in SE

A recent research trend has emerged to leverage sentiment analysis for enhancing software development, maintenance, and evolution by applying sentiment analysis on Stack Overflow [32], app reviews [21], and tweets containing comments about software applications [14]. Further research investigates the role of affect in social software engineering, by applying sentiment analysis to the content available in collaborative development environments such as GitHub [15][31][38], Jira [22][27], and Stack Overflow [7]. With a notable few exceptions [3][27][30], empirical software engineering studies have exploited off-the-shelf sentiment analysis tools trained on non-software engineering documents. However, there is a consensus in the research community about the negative results obtained when using such tools to detect developers' emotions [16][20][26].

Trying to overcome the limitations posed by off-the-shelf sentiment analysis tools, some researchers recently started to develop their own SE-specific tools. Panichella et al. [30] trained their own classifier on 2,000 manually-annotated reviews in Google Play and Apple Store, using a bag-of-word approach and Naïve Bayes for training. Blaz and Becker [3] developed a polarity classifier for IT tickets. They implemented an approach based on a domain dictionary created using semiautomatic bootstrapping to expand an initial set of affectively-loaded words used as seeds. They compare different feature settings obtaining, in the best experimental setting, an overall performance of F = .85. However, their classifier still reports a negative bias inducing the misclassification of neutral documents as negative. Other dimensions of affect have also been studied. It is the case of Mäntylä et al. [23], who bootstrapped an SE-specific lexicon for arousal, that is the level of emotional activation ranging from low (relaxed) to high (excited). Gachechiladze et al. [12] reported about a preliminary investigation of supervised approaches to detection of the anger target in developers' communication in Jira.

## 3. SENTIMENT ANALYSIS TOOLS

At the time of writing, three SE-specific sentiment analysis tools are publicly available, namely Senti4SD [5], SentiStrengthSE [18], and SentiCR [1]. We include them to address RQ1 and RQ2 and provide detailed information on their design, implementation, and usage in this section. Our baseline is represented by SentiStrength, the most widely employed tool in sentiment analysis studies in software engineering [4][7][15][21][38].

### 3.1 SentiStrength (baseline)

SentiStrength [40] is a state-of-the-art, lexicon-based classifier, which can deal with short informal text, including abbreviations, intensifiers, and emoticons. In the SentiStrength lexicon, each negative word receives a sentiment score ranging from −2 to −5, which represents its *a priori* polarity (i.e., the polarity of the term out of its contextual use). Similarly, positive words are associated with a score between +2 and +5, while neutral words receive scores equal to ±1. Positive and negative emoticons are also included in the dictionary. Grounded on the assumption that a sentence can convey mixed sentiment, SentiStrength outputs both positive and negative sentiment scores for an input text. Given an input sentence, SentiStrength determines its overall positive and negative scores by choosing the maximum positive and negative scores of the terms in the sentence. Intensifiers, i.e., exclamation marks or verbs such as 'really', are treated as booster words and increase the a priori word sentiment scores. The polarity scores of words in the scope of negations are inverted. Similarly, if the input document includes multiple sentences, the maximum values of all the sentence scores are considered. Positive sentiment scores range from + 1 (absence of positive sentiment) to + 5 (extremely positive) while negative sentiment scores range from − 1 (absence of negative sentiment) to − 5 (extremely negative). Based on their algebraic sum, SentiStrength can also report the overall trinary score, i.e. the overall positive (score = 1), negative (score = -1) or neutral (score = 0). Consistently with the study, we are currently replicating [16], we map SentiStrength scores using this trinary approach.

### 3.2 Senti4SD

Senti4SD is a polarity classifier specifically trained to support sentiment analysis in developers' communication channels [5]. Senti4SD is publicly available for research purposes and is distributed together with a classification model trained and validated on a gold standard of about 4K questions, answers, and comments extracted from Stack Overflow and manually annotated for sentiment polarity. Furthermore, the toolkit provides a training method that enables the training of custom classifiers from scratch, using an annotated gold standard as input. Senti4SD leverages a suite of features based on n-grams, sentiment lexicons and semantic features based on word embedding, whose contribution is assessed by the authors through an empirical evaluation leveraging different feature settings. With respect to the performance obtained by SentiStrength on the same Stack Overflow dataset, Senti4SD reduces the misclassifications of neutral and positive posts as emotionally negative. Senti4SD achieves a comparable performance also in presence of a minimal set of training documents.

### 3.3 SentiStrengthSE

SentiStrengthSE is an SE-specific sentiment analysis tool, built upon the SentiStrength API [18]. It leverages a manually adjusted version of the SentiStrength lexicon and implements ad hoc heuristics to correct the misclassifications observed when running SentiStrength on the Ortu et al. dataset [28]. The sentiment scores of words in the lexicon were manually adjusted to reflect the semantics and neutral polarity of domain words such as 'support', 'error', or 'default'. The evaluation performed by the authors shows that SentiStrengthSE outperforms SentiStrength on technical texts.

### 3.4 SentiCR

SentiCR is a supervised sentiment analysis toolkit, specifically trained and evaluated for code review comments [1]. SentiCR leverages a feature vector generated by computing TF-IDF (Term Frequency - Inverse Document Frequency) for bag-of-words extracted from the input text. Before generating the feature vector, SentiCR implements raw text preprocessing to normalize text (i.e., expansion of contractions as in 'don't' or 'I'm'), handle negations and emoticons, remove stop-words (i.e., articles, prepositions, conjunctions, and pronoun), derive word stems and remove code snippets. Furthermore, it performs SMOTE (synthetic minority over-sampling technique) [8] to address the problem of class imbalance in the training data. SentiCR has been evaluated using eight supervised algorithms in a 10-fold cross validation setting. The currently distributed version of SentiCR implements a training approach based on Gradient Boosting Tree.

## 4. METHODOLOGY

In this Section, we describe the evaluation metrics (Sec. 4.1), the datasets included in the benchmark (4.2), and the experimental setting adopted for assessing the tools' performance (4.3).

### 4.1 Evaluation Metrics

We report the performance of each sentiment analysis tool in terms of precision, recall, and f-measure for all the three polarity classes. This choice is in line with the standard methodology adopted for benchmarking of sentiment analysis systems in evaluation campaigns [2][25] and, more in general, for assessing text categorization approaches [35]. Precision is the ratio between the true positive and all the predicted items for a given polarity class. Recall represents the ratio of true positives and all items belonging to a given polarity class. F-measure is computed as the harmonic mean of precision and recall. Additionally, we report micro- and macro-averaged values for precision, recall, and f-measures to enable quick comparison of the overall performance of each classifier. The overall performance is computed adopting micro-averaging as aggregated metric. Given the unbalanced distribution of labels in three of the four datasets in our benchmark, we also report macro-average, i.e., precision and recall are first evaluated locally for each class, and then globally by averaging the results of the different categories. Indeed, micro- and macro- averaging may lead to different results. For instance, the performance on classes with few positive training instances is emphasized by macro-averaging. Conversely, micro-averaging tends to be mainly influenced by the performance on the majority class. Whether to optimize by precision, recall, or f-measure obviously depends on the application scenario, as discussed by Calefato et al. [5].

Furthermore, we report the weighted kappa ($\kappa$) by Cohen [9], a measure of interrater agreement that we use to assess both the agreement with gold labels (RQ1) and the agreement among the three tools (RQ2). We are interested in distinguishing between *mild disagreement*, that is the disagreement between negative/positive and neutral annotations, and *strong disagreement*, that is the disagreement between positive and negative judgments. As such, we assigned a weight = 2 to strong disagreement and a weight = 1 to mild disagreement in computing the weighted Kappa (see Table 1). We follow the interpretation of $\kappa$ by Viera and Garrett [41], suggesting that the agreement is less than chance if $\kappa \leq 0$, slight if $0.01 \leq \kappa \leq 0.20$, fair if $0.21 \leq \kappa \leq 0.40$, moderate if $0.41 \leq \kappa \leq 0.60$, substantial if $0.61 \leq \kappa \leq 0.80$ and almost perfect if $0.81 \leq \kappa \leq 1$. Both, the weighted scheme and the interpretation of $\kappa$ are the same adopted in the original study by Jongeling et al. [16].

### 4.2 Datasets

In our benchmark, we include four publicly available datasets specifically developed to investigate the role of sentiment in software development. In Table 2, we report the overall number of items and the class distribution for each set. More details about the model-driven and ad hoc annotation approaches, adopted for each dataset, are provided in the following.

**Model-driven annotation.** Two datasets have been annotated following a model-driven approach, namely the Stack Overflow dataset [5] and the Jira dataset [28]. Both datasets were labeled according to the Shaver et al. framework [37], a tree-structured hierarchical classification of emotions. The model organizes emotion labels in three levels of hierarchical clusters. Each level refines the granularity of the previous one, thus providing more indication on its nature. The framework includes, at the top level, six basic emotions, namely love, joy, anger, sadness, fear, and surprise. It is easy to understand, thanks to the intuitive nature of the emotion labels and their direct mapping to polarity.

The **Stack Overflow** dataset already implements this mapping with polarity labels for 4,423 posts, including questions, answers, and comments. The dataset is well-balanced: 35% of posts convey positive emotions while 27% present negative emotions. No emotions are observed for the remaining 38% of posts, thus they receive the neutral polarity label. The dataset has been annotated by twelve trained coders with a background in Computer Science. Coders were requested to explicitly indicate a polarity label for each post according to the specific affect detected. Each post was annotated by three raters and received the polarity gold label based on majority voting. As an evidence of the reliability of the annotation schema, authors report an average weighted Cohen's Kappa of .74.

The **Jira** dataset includes about 2,000 issue comments and 4,000 sentences contributed by developers using Jira. Specifically, Jira-based issue comments come from popular open source software projects, such as Apache, Spring, JBoss, and CodeHaus. The Jira dataset is distributed with emotion labels from the Shaver framework. It includes three subsets: group 1 is composed of 392 comments labelled with emotions love, joy, surprise, anger, fear and sadness; group 2 contains 1,600 comments labelled with emotions love, joy, sadness; group 3 includes 4,000 sentences, extracted from issue comments, labelled with emotions love, joy, anger, and sadness. For groups 1 and 2, the original annotations by all raters are distributed. As such, we assigned the gold emotion label for each item based on majority voting. Conversely, for group 3 the computation of majority voting was not necessary as the authors directly provide the gold emotion label for each text item. We assigned the final polarity label for all documents in the Jira dataset by directly mapping the gold emotion labels into the corresponding polarity classes. Consistently to the approach used for the Stack Overflow dataset [5] as well as in the original study we are replicating [16], we represent joy and love as positive polarity; sadness, fear, and anger as negative polarity; and absence of emotions as neutral polarity. Surprise is the only emotion that could match either positive or negative polarity. In the annotation of the Stack Overflow dataset, raters were required to disambiguate polarity of surprise based on context information. However, this is not possible for the Jira dataset as this would involve making assumptions on the original annotation. To avoid introducing any bias or noise, we decided to discard the Jira comments originally annotated as surprise.

**Table 1. Weighting scheme for computation of weighted $\kappa$.**

|  | Negative | Neutral | Positive |
|---|---|---|---|
| Negative | 0 | 1 | 2 |
| Neutral | 1 | 0 | 1 |
| Positive | 2 | 1 | 0 |

**Table 2. Datasets included in our benchmark with distribution of polarity classes.**

| Dataset | Overall Items | Unit of Annotation | Polarity Classes and Distribution | | |
|---|---|---|---|---|---|
| | | | *Neg* | *Pos* | *Neu* |
| *Model-driven annotation* | | | | | |
| Stack Overflow | 4423 | Post | 27% | 35% | 38% |
| Jira | 5869 | Sentence/entire comment | 13% | 19% | 68% |
| *Ad hoc annotation* | | | | | |
| Java Libraries | 1500 | Sentence | 12% | 9% | 79% |
| | | | *Neg* | *Non-neg* | |
| Code Review | 1600 | Comment | 25% | 75% | |

**Ad hoc annotation.** The Code Review and Java Libraries corpora have been annotated following an ad hoc approach, i.e., by asking the raters to annotate the positive, negative, or neutral semantic orientation of a text based on their subjective perception without any further guidance. The **Code Review** corpus is the dataset used for the original training and evaluation of SentiCR [1]. It includes 2000 review comments from 20 popular open source software projects using Gerrit. Three researchers annotated the comments as 'positive', 'negative' or 'neutral', based on what they would perceive if they were the comment recipients. Ahmed et al. [1] motivate the choice to adopt an ad hoc approach with their intention to capture a true sentiment perception. A moderate agreement is observed, as authors report 62.5% of perfect agreement and a κ = 0.408. To assign the gold labels, disagreements were solved through discussion. Furthermore, to address the problem of class imbalance, authors decided to merge the positive and neutral classes using a non-negative label.

The **Java Libraries** dataset was collected in the scope of a broader study aiming at developing a recommender for software libraries that leverages sentiment analysis for mining crowdsourced opinions [20]. The dataset includes 1,500 sentences, randomly extracted from Stack Overflow. Five researchers manually labeled sentences by assigning a sentiment score to each sentence. The assessment of interrater agreement in terms of κ is not provided. However, authors report that disagreement occurred in about 18% of cases.

### 4.3 Experimental setting

In our study, we compare two supervised tools, namely Senti4SD and SentiCR, and one unsupervised classifier, namely SentiStrengthSE. To enable a fair comparison, we split each gold set into training (70%) and test (30%) sets, using the R package caret [17] for stratified sampling. We used the training set to retrain the supervised classifiers, by replicating the experimental setting described in the original studies, using the training methods made available in both Senti4SD and SentiCR toolkits. The performance of each tool is then assessed on the test set, which is also used to build the SentiStrength baseline. Both SentiStrength and SentiStrength-SE are implemented by leveraging a lexicon-based approach (see Sections 3.1 and 3.3) so they are not retrained.

## 5. RESULTS

In this Section, we report the results of our empirical study with respect to our research questions.

### 5.1 Performance of Sentiment Analysis Tools

*RQ1: To what extent do different SE-specific sentiment analysis tools agree with emotions of software developers?*

In Table 3, we report the performance obtained in terms of recall, precision, and F-measure, for each polarity classes as well as the overall performance, for the three tools on the model-driven annotation datasets. In bold we highlight the best values for each metric. Furthermore, in Table 4(a) we report their agreement with manual labeling, in terms of weighted Cohen κ. In addition to interrater agreement, we report the percentage of cases for which the tool issues the correct prediction (*Perfect agreement* with gold label) as well as the percentage of cases for which *Severe* (positive vs. negative, and vice versa) and *Mild Disagreement* (positive/negative vs. neutral) is observed.

We observe that for both the Stack Overflow and Jira datasets, the SE-specific tools outperform the SentiStrength baseline. This evidence suggests that the customization of sentiment analysis tools to the software development domain does improve the classification accuracy. The improvement is higher for the Jira dataset, for which we observe the best performance using SentiCR, probably due to the SMOTE optimization of label distribution in the training set (F-measure micro = .85 and macro = .80, κ = .73). Similar performance is observed for SentiStrengthSE (F-measure micro = .83 and macro = .78, κ = .70) and Senti4SD (F-measure micro = .83 and macro = .76, κ = .67). Conversely, the baseline approach on Jira produces an overall performance lower than .70 (F micro = .69 and macro = .65, κ = .48). As for the Stack Overflow dataset, the best performing approaches is Senti4SD (F-measure micro = .87 and macro = .86, κ = .83), followed by SentiCR (F-measure micro = .82 and macro = .82, κ = .76) and SentiStrengthSE (F-measure micro = .80 and macro = .80, κ = .74). The good performance of SentiStrength on the Stack Overflow dataset can be explained by the fact that the annotation corpus for this dataset was created by performing opportunistic sampling of posts based on the presence of affective terms belonging to the SentiStrength lexicon [5].

The higher recall for negative class and precision for neutral class observed for SentiStrength on the Stack Overflow dataset confirms its bias towards the recognition of negative sentiments, already highlighted by previous research [5]. In fact, SentiStrength erroneously classifies many neutral posts as negative, with a poor recall for neutral class (.64) and a low precision for the negative one (.67). All SE-specific tools are able to correct the problem of such negative bias. The best performance is observed for Senti4SD (F-measure micro = .87 and macro = .86, κ = .83): the F-measure increases from .79 (with SentiStrength) to .84 (with Senti4SD) for the negative class and from .76 to .83 for the neutral class, thus depicting a more balanced classifier (see Table 3). This is the consequence of an improved recall of neutral documents from .64 up to .80 (25% of improvement) and the precision of negative documents from .67 up to .80 (19% of improvement).

Table 3. Performance of sentiment analysis tools for model-driven annotations.

| Dataset | Class | SentiStrength (baseline) | | | Senti4SD | | | SentiStrengthSE | | | SentiCR | | |
|---|---|---|---|---|---|---|---|---|---|---|---|---|---|
| | | P | R | F1 | P | R | F1 | P | R | F1 | P | R | F1 |
| Stack Overflow | Positive | .89 | **.92** | .90 | **.92** | **.92** | **.92** | .89 | .83 | .86 | .88 | .90 | .89 |
| | Negative | .67 | **.96** | .79 | **.80** | .89 | **.84** | .75 | .79 | .77 | .79 | .73 | .76 |
| | Neutral | **.95** | .64 | .76 | .87 | .80 | **.83** | .75 | .77 | .76 | .79 | **.82** | .80 |
| | Micro-avg. | .82 | .82 | .82 | **.87** | **.87** | **.87** | .80 | .80 | .80 | .82 | .82 | .82 |
| | Macro-avg. | .84 | .84 | .84 | **.86** | **.87** | **.86** | .80 | .80 | .80 | .82 | .81 | .82 |
| Jira | Positive | .50 | **.91** | .65 | **.76** | .79 | .78 | .69 | .94 | .80 | **.76** | .89 | **.82** |
| | Negative | .41 | .64 | .50 | .72 | .57 | .64 | .67 | .71 | .69 | **.81** | .61 | **.70** |
| | Neutral | .89 | .59 | .71 | .86 | **.89** | .88 | **.92** | .82 | .87 | .89 | **.89** | **.89** |
| | Micro-avg. | .66 | .66 | .66 | .83 | .83 | .83 | .83 | .83 | .83 | **.85** | **.85** | **.85** |
| | Macro-avg. | .60 | .71 | .62 | .78 | .75 | .76 | .76 | **.82** | .78 | **.82** | .80 | **.80** |

Table 4. Agreement of SE-specific tools with manual labelling (a) and with each other (b) for model-driven annotations.

| | | (a) | | | | (b) | | | | |
|---|---|---|---|---|---|---|---|---|---|---|
| Dataset | Classifier | Agreement metrics | | | | Classifiers | Agreement metrics | | | |
| | | k | Perfect Agreem. | Disagreement Severe | Mild | | k | Perfect Agreem. | Disagreement Severe | Mild |
| Stack Overflow | Senti4SD | .83 | 86% | 1% | 12% | Senti4SD vs. SentiCR | .77 | 83% | 3% | 14% |
| | SentiStrengthSE | .74 | 80% | 2% | 18% | Senti4SD vs. SentiStrengthSE | .79 | 84% | 2% | 15% |
| | SentiCR | .76 | 82% | 3% | 15% | SentiCR v. SentiStrengthSE | .73 | 80% | 3% | 17% |
| | *SentiStrength (baseline)* | .77 | 82% | 3% | 15% | | | | | |
| Jira | Senti4SD | .67 | 83% | 0 | 17% | Senti4SD vs. SentiCR | .76 | 88% | 0 | 12% |
| | SentiStrengthSE | .70 | 83% | 0 | 17% | Senti4SD vs. SentiStrengthSE | .70 | 83% | <1% | 16% |
| | SentiCR | .73 | 86% | 0 | 14% | SentiCR v. SentiStrengthSE | .81 | 89% | <1% | 10% |
| | *SentiStrength (baseline)* | .48 | 66% | <2% | 33% | | | | | |

Misclassification of neutral texts as either positive or negative also occurs when using SentiStrength on the Jira dataset. Again, all SE-specific tools are able to correct this bias. In particular, the best performance is observed for SentiCR (F-measure micro = .85 and macro = .80, κ = .73): the F-measure increases from .50 (SentiStrength) to .70 (SentiCR) for the negative class, from .65 to .82 for the positive class, and from .71 to .89 for the neutral class. This is the consequence of an improved recall of neutral documents from .59 up to .89 (51% of improvement) and the precision of negative and positive comments from .41 up to .81 (97% of improvement) and from .50 up to .76 (52% of improvement), respectively.

Based on the evidence provided by the agreement metrics in Table 4(a), we observe that the disagreement between tool predictions and gold labels is mainly between positive/negative and neutral (*mild disagreement*). *Strong disagreement*, i.e. between positive and negative classes, never occurs for the Jira dataset and is observed only in 1% of cases for the best performing tool on the Stack Overflow dataset.

*RQ2: To what extent do results from different SE-specific sentiment analysis tools agree with each other?*

In Table 4(b), we report the agreement between tools on the model-driven annotation datasets. Once again, in addition to weighted Cohen κ, we assess the interrater agreement also in terms of percentage of cases for which the tools agree (*perfect agreement*) as well as the percentage of cases for which *severe* (positive vs. negative, and vice versa) and *mild disagreement* (positive/negative vs. neutral) occur. We observe substantial to perfect agreement for all couples of tools. Again, *mild disagreement* is the main cause of the difference in performance. *Strong disagreement* is never observed on the Jira dataset and is equal to 3% in the worst case on Stack Overflow.

*RQ3: To what extent do the labeling approach (model-driven vs. ad hoc annotation) has an impact on the performance of SE-specific sentiment analysis tools?*

We replicate the performance assessment on ad hoc annotation datasets, namely Code Review and Jira Libraries. Results are reported in Tables 5 and 6. Once again, we report recall, precision, and F-measure (see Table 5) as well as the assessment of agreement (see Table 6).

Compared to the model-driven annotation, we observe a drop in performance for both datasets. In fact, even if the F-measure micro averages are comparable to those observed in model-driven dataset, F-measure macro averages indicate lower performance due to class imbalance. Specifically, we observe better performance on the neutral class, which is the majority class for both datasets. Lower performance on the ad hoc annotation datasets is also reflected in values of κ ranging from slight to moderate agreement, with the highest values for κ = .50 observed for SentiCR on the Code Review comments. Still, *strong disagreement* is rarely observed (2% in the worst case).

## 5.2 Error Analysis

To get a deeper insight on the difficulties inherent to sentiment detection in software engineering, we manually examined cases for which all the three tools yielded a wrong prediction. The goal of this analysis is twofold: on one hand, we aim at identifying open challenges in sentiment analysis of developers' communication traces, even in presence of good performance, as observed for model-driven annotation dataset; on the other hand, we aim at assessing to what extent misclassification is induced by the adoption of an ad hoc approach to annotation. In the following, we report and discuss notable error classes resulting from our analysis of misclassified texts, ordered by frequency of observation. The overall number of items misclassified by all tools is reported in Table 7, with breakdown by dataset.

Each text in the misclassified set was annotated by one of the authors with possible causes of errors. The results were collectively discussed to identify potential reasons and error patterns. Please, note that we used and applied specific categories to better understand the systems' shortcomings, and we report examples for each error category. However, it may be the case that more than one error category applies to single documents and it is the co-occurrence of difficulties that can be responsible for misclassification. The overall distribution of error categories, as well as their distribution in each dataset, are reported in Figures 1 and 2, respectively.

**Polar facts but neutral sentiment.** - Among misclassified instances, there are comments or posts that involve a `polar fact' [42], that is information, such as a problem report or the acceptance of a patch, where the reported event usually invokes for most people a positive or negative feeling, i.e. the annotator considered the described situation either as desirable or undesirable. See for instance the following comments reporting a problem with the code they refer to:

*'This creates an unnecessary garbage list. Sets.newHashSet should accept an Iterable.'*

or

*'I tried the following and it returns nothing'.*

Both comments are neutral as they not conveyed any emotions nor opinion. Still, they are both annotated as negative as they involve description of polar facts. This is the main cause for misclassification for ad hoc annotation datasets, with 61% and 89% of misclassified texts for Code Review and Java Libraries, respectively. Conversely, we observe this error only in 7% and 12% of misclassified texts for Stack Overflow and Jira datasets, respectively (see Figure 2). This result suggests that model-driven annotation enhances the reliability of gold standard datasets for sentiment analysis, as it supports annotators in disambiguating between actual emotional comments and neutral reports of bugs and fixes.

**Table 5. Performance of sentiment analysis tools for ad hoc annotations.**

| Dataset | Class | SentiStrength (baseline) | | | Senti4SD | | | SentiStrengthSE | | | SentiCR | | |
|---|---|---|---|---|---|---|---|---|---|---|---|---|---|
| | | P | R | F1 | P | R | F1 | P | R | F1 | P | R | F1 |
| Code Review | Negative | .36 | .33 | .35 | **.68** | .40 | .51 | .53 | .25 | .34 | .64 | **.64** | **.64** |
| | Non-Negative | .79 | .81 | .80 | .83 | **.94** | **.88** | .79 | .93 | .85 | **.88** | .88 | .88 |
| | Micro-avg. | .69 | .69 | .69 | .80 | .80 | .80 | .76 | .76 | .76 | **.82** | **.82** | **.82** |
| | Macro-avg. | .58 | .57 | .57 | .75 | .70 | .69 | .66 | .59 | .60 | **.76** | **.76** | **.76** |
| Java Libraries | Positive | .11 | .23 | .15 | **.63** | .26 | .36 | .19 | .15 | .17 | .50 | **.33** | **.40** |
| | Negative | .34 | .31 | .33 | **.55** | .33 | .41 | .40 | .13 | .19 | .49 | **.52** | **.50** |
| | Neutral | .81 | .73 | .77 | .84 | **.96** | **.90** | .80 | .92 | .85 | **.87** | .89 | .88 |
| | Micro-avg. | .62 | .62 | .62 | **.80** | **.80** | **.80** | .74 | .74 | .74 | .79 | .79 | .79 |
| | Macro- avg. | .42 | .42 | .41 | **.67** | .51 | .56 | .46 | .40 | .40 | .62 | **.58** | **.59** |

**Table 6. Agreement of SE-specific tools with manual labelling (a) and with each other (b) for ad hoc annotations.**

| | | (a) | | | | | (b) | | | |
|---|---|---|---|---|---|---|---|---|---|---|
| Dataset | Classifier | Agreement metrics | | | | Classifiers | Agreement metrics | | | |
| | | k | Perfect Agreem. | Disagreement | | | k | Perfect Agreem. | Disagreement | |
| | | | | Severe | Mild | | | | Severe | Mild |
| Code Review | Senti4SD | .34 | 80% | NA | 20% | Senti4SD vs. SentiCR | .40 | 81% | NA | 19% |
| | SentiStrengthSE | .28 | 76% | NA | 24% | Senti4SD vs. SentiStrengthSE | .26 | 83% | NA | 17% |
| | SentiCR | .52 | 82% | NA | 18% | SentiCR v. SentiStrengthSE | .20 | 76% | NA | 24% |
| | *SentiStrength (baseline)* | .21 | 76% | NA | 24% | | | | | |
| Java Libraries | Senti4SD | .34 | 81% | 2% | 17% | Senti4SD vs. SentiCR | .48 | 85% | 0 | 15% |
| | SentiStrengthSE | .12 | 74% | 2% | 24% | Senti4SD vs. SentiStrengthSE | .16 | 82% | 1% | 17% |
| | SentiCR | .40 | 79% | 2% | 19% | SentiCR v. SentiStrengthSE | .16 | 77% | 2% | 22% |
| | *SentiStrength (baseline)* | .13 | 62% | 4% | 34% | | | | | |

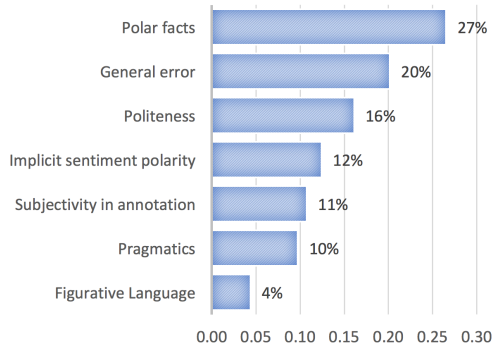

**Figure 1. Distribution of error categories**

Table 7. Distribution of texts misclassified by all tools.

|  | Misclassified (% of the test set) | Test set size |
|---|---|---|
| Stack Overflow | 88 (7%) | 1.326 |
| Jira | 101 (6%) | 1.759 |
| Code Review | 33 (7%) | 480 |
| Java Libraries | 50 (11%) | 449 |
| Overall | 272 (7%) | 4.014 |

**General Error.** These errors are due to the tool inability to deal with some textual cues or errors in preprocessing raw text. In some cases, lexical cues are not recognized because they are not included in the sentiment dictionary and are probably not frequent enough to have sufficient predictive power (e.g., '+1' in *'+1 for not being afraid to ask'*). Another error is wrong negation handling as in

*'Could you add some details about the error you get? Nothing is jumping out at me as horribly wrong.',*

sometimes due to broken syntax, as in *'wontbe so bad'*.

Also, idiomatic expression should be considered to yield the correct polarity label, as in *'Are you out of your mind?'*.

Finally, some long posts are classified as neutral probably due to the prevalence of neutral lexicon, even in presence of emotion words as in

*'Yes, if I were to declare the property and synthesize it as above, but I don't explicitly declare an ivar myself, the compiler will still actually create the ivar. So I could do `int len = [self.stringProperty length]` as you'd expect inside the class, AND I can also do `int len = [stringProperty length]` as well, without using the accessor method, just directly using the ivar (which the compiler has synthesized for me, along with setter+getter). It's really fantastic!*

This suggests that sentence-based annotation might enhance the performance of sentiment classification when long posts are analyzed.

**Politeness.** Depending on the context, raters may provide polar or neutral labels to sentences containing lexical cues of politeness. It is the case, for example, of texts including lexical forms like 'Thanks' or 'Thank you'. Such texts might receive a neutral label when politeness is perceived by the rater as in

*'Jimmy this ready for review/commit? If so please 'submit'. Thanks.'*

Conversely, a positive label is assigned when the raters perceive the speaker's intention to communicate actual emotions beyond politeness, such as gratitude or appreciation as in

*'works excellent!! thanks a lot guys!!!'.*

Similarly, use of negative lexicon as 'Sorry' or 'I'm afraid' can be perceived either as a simple sign of politeness, as in

*'While this could work when I'm developing and debugging the solution, I'm afraid it wouldn't work once the solution is deployed and a user needs to change the settings')*

or as a cue of the presence of negative affective states such as sadness, as in

*'yes it was a horrible syntax error, sorry guy's!'.*

In presence of such ambiguous lexical features, classifiers are not able to successfully distinguish between politeness (neutral sentiment) and presence of actual emotions, unless the presence of other lexical clues enables disambiguation. Based on this evidence, we envisage the need of considering the politeness scores [10] in sentiment polarity classification, as already done in similar studies on emotion recognition [6][27].

**Implicit sentiment polarity.** In 12% of cases overall (see Figure 1), the tools fail to detect either positive or negative sentiment due to the absence of explicitly polarized lexical clues. In these cases, developers use instead neutral lexicon to describe emotion-triggering situations or attitudes towards the interlocutor. However, a correct evaluation of such text can nonetheless be inferred by human annotators by relying on commonsense knowledge or contextual information. It is the case, for example, of anger towards others (i.e., peers, colleagues) expressed using neutral lexicon, which is a possible consequence of the 'emotional labor' of developers trying to comply to code of conduct adopted in the community [36]. For example, human can perceive and correctly annotate the aggressive attitude in

*'I don't have to ensure that the classloader knows groovy classes \*you\* must do that'*

or in *'Please accept that my l'ogical"is different from your l'ogical"*

where the use of punctuation is a subtle sign of hostile attitude towards the interlocutor. The inability to correctly recognize such cases as negative confirms the difficulty in recognizing the anger towards others described by previous research on anger detection [12]. Similarly, individual emotions, as frustration, can be conveyed by describing undesired situation triggering negative sentiment with neutral lexicon, as in *'25 hours gone'*.

Similar cases are observed for positive sentiment, as in the comment

*'I think we are very close too'*

in which the author uses a neutral lexicon to convey the optimism for being close to a solution to his problem.

**Subjectivity in sentiment annotation.** The annotation of sentiment is known to be a subjective task, as the perception of emotions might vary according to personality traits or personal disposition [34]. In some cases, raters were too conservative in interpreting the guidelines for annotation and provided a neutral majority label even in presence of positive or negative emotions or opinions.

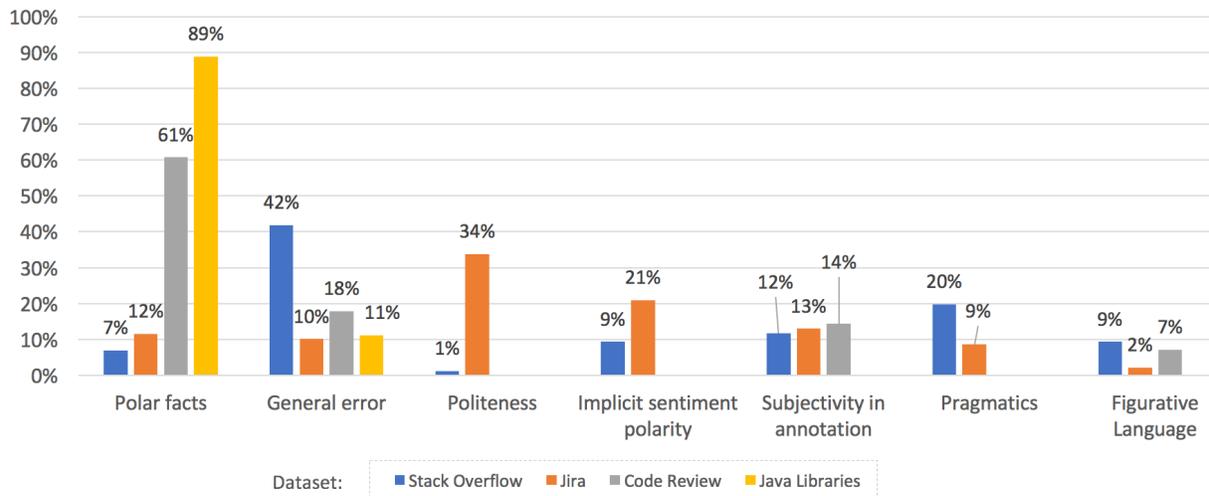

**Figure 2. Distribution of error categories.**

**Inability of classifiers to deal with pragmatics or context information.** Human raters consider some texts neutral even in presence of lexical clues of affect or opinions because they can disambiguate the semantic orientation of the text based on the actual communicative intention of the speaker.

It is the case, for example of the description of hypothetical scenarios, e.g.,

*'Something that would at least avoid the pain of would be <code>'*

report of another's emotions or opinions as in

*'Side note: some people hate that menu.'*

or questions, as in

*'Just wondering; do you hate the string variable type?'*

Similarly, we observe that ten neutral comments from the Jira dataset were misclassified as positive because they report a positive polar fact, as in *'The patch looks good'*.

Cases like this are extremely challenging for a text-based classifier if the system does not embed any rule or strategy to deal with the pragmatic context, such as recognition of speech acts (i.e., questions, statements, etc.) or with representation of domain knowledge about neutral polar facts that are frequent in a specific context of interaction (i.e., approving a patch or reporting a bug).

**Figurative language.** The use of figurative language to convey humor, irony, and sarcasm, often expressed through neutral lexicon, as in

*'Maybe only one of us should play lawyer and the other one play engineer.',*

is another cause of misclassification. Indeed, irony detection and sentiment analysis of figurative language are complex tasks, that received dedicated attention in recent evaluation campaigns for natural language tasks [2][13].

## 6. DISCUSSION

In this paper, we describe a partial replication of the benchmarking study performed by Jongeling et al. [16]. The original study showed that not only off-the-shelf sentiment analysis tools disagree with the emotions annotated in developers' communication traces but they also disagree among themselves. A more recent study by Lin et al. [20] confirm these results and concludes that no tool is ready for real usage in the software engineering domain.

We contrast this evidence by showing that SE-specific tuning may improve the accuracy of sentiment analysis, provided that theoretical models of affect are taken into due consideration. Our results indicate that not only SE-specific tools obtain good agreement with respect to manual labeling (R1) but they also agree with each other (RQ2). Also, the substantial agreement between tools should warn us about the presence of common problems that still represent open challenges for sentiment analysis in software engineering, even in presence of the satisfying performance observed for model-driven annotation (see Tables 3 and 4). In search of an explanation, in the previous Section we also report the results of an error analysis on the documents that were misclassified by all tools. In the following, we summarize the lessons learned by combining evidence from both quantitative and qualitative analysis, to inform future research on sentiment analysis in software engineering.

**Lesson: Reliable sentiment analysis in software engineering is possible.** We provide empirical evidence that sentiment analysis can be reliably performed on developers' communication traces. Specifically, we show that SE-specific tuning of tools improves accuracy with respect to the SentiStrength baseline.

**Lesson: Tuning of tools in the software engineering domain enhances accuracy.** SE-specific customization of sentiment analysis tools produces better classification accuracy. Improved performance is observed with respect to results reported in previous studies assessing off-the-shelf tools on software engineering datasets [16][20]. Supervised tools, namely Senti4SD and SentiCR, outperform SentiStrengthSE, which implements a lexicon-based approach. However, as also observed by Lin et al. [20], the customization might produce a different performance on different data sources. As such, we recommend retraining supervised tools by leveraging a gold standard from the same domain and data source object of study, as lexical choice and sentiment interpretation may vary depending on the specific communication style of the platform considered. All supervised tools analyzed enable retraining based on custom gold standards. Specifically, Calefato et al. [5] provided evidence that Senti4SD produces optimal performance even with a minimal set of training documents.

**Lesson: Preliminary sanity check is always recommended**. Whenever re-training of supervised classifiers is not possible due to the unavailability of a manually annotated gold standard, the unsupervised approach implemented by SentiStrengthSE provides comparable performance. However, we should be aware that tools and datasets are built by having in mind specific research goals and different conceptualization of affect [26]. As such, a sanity check is always recommended to assess the suitability of existing tools with respect to our research goals, to avoid introducing threats to validity due to erroneous classification of emotions and opinions.

**Lesson: Grounding research on theoretical models of affect is recommended.** As discussed above, re-training of tools in the software engineering domain is beneficial for performance. Comparison of performance between model-driven and ad hoc annotated datasets shows that retraining might still be not enough to guarantee a satisfactory accuracy for all polarity classes if ad hoc annotation is adopted for building the gold standard. This is consistent with previous research, suggesting that the choice of the model underlying operationalization of emotions might affect the construct validity of empirical studies [26]. Furthermore, given the wide variety of affective states expressed by developers [7][24], we advocate in favor of a more fine-grained investigation of the role of emotions, in line with recent research [6][12]. In fact, depending on the specific goals addressed, researchers could be interested in detecting specific emotions (e.g., frustration, anger, sadness, joy, satisfaction), opinions (i.e., positive or negative evaluations), or interpersonal stances (i.e., friendly vs. hostile attitude). We also showed how the absence of clear guidelines for annotation leads to noisy gold standards where neutral polar facts, such as bug reports or approval of patches, are erroneously labeled as negative or positive by the human raters. Thus, we underline the need to distinguish between the task of identifying the affective content conveyed by a text (i.e., sentiment analysis) and the task of identifying the objective report of (un)pleasant facts in the developers' comments.

## 7. THREATS TO VALIDITY

Conclusion validity is influenced by the choice of datasets to include in our benchmark, as our methodology could produce different results if applied to different datasets. However, we included all the publicly available gold standards for sentiment annotation in software engineering, available at the time of writing. All datasets are built by collecting text from popular platforms among software developers. As such, we are reasonably confident that the datasets we have considered are representative of developers' communication traces. Nevertheless, we acknowledge that replications are needed to further increase the external validity to the entire software developers' ecosystem.

The reliability of benchmarking in sentiment analysis research is affected by the inherent subjectivity of the phenomenon being studied, i.e., emotions and opinions as expressed in written communication. Indeed, emotion annotation is a subjective process since affect triggering and perception can be influenced by personality traits and personal dispositions [34]. This threat is not only mitigated but is explicitly investigated in our third research question, by comparing the performance of tools on model-driven and ad hoc annotation datasets.

Finally, threats to internal validity concern internal factors such as the configuration of the parameters for the machine learning algorithms implemented by Senti4SD and SentiCR. To mitigate this threat, we replicated the experimental conditions under which both tools were originally validated by Calefato et al. [5] and Ahmed et al. [1]. Specifically, fine-tuning of C parameter based on accuracy was performed before choosing the final model for Senti4SD. As for SentiCR, SMOTE-based optimization is embedded in the training function released by the authors. Also, retraining was performed to enable a fair comparison of supervised classifiers. However, we acknowledge it could be useful to compare the performance against the original classifiers, as tools can be also used out of the box if a gold standard for retraining is not available.

## 8. CONCLUSION

We have reported a benchmarking study aimed at assessing the performance of sentiment analysis tools specifically tuned for the software engineering domain. We included in our benchmark four gold standard datasets including developers' communication traces from channels of the software developers' ecosystem. We found that reliable sentiment analysis in software engineering is possible, provided that manual annotation of gold standards is inspired by theoretical models of affect. Regardless of the approach adopted for annotation, SE-specific customization does provide a boost in accuracy with respect to the baseline approach represented by an off-the-shelf tool. Specifically, the best performance is observed for supervised approaches thus leading to our recommendation of performing custom retraining of classifiers. However, whenever retraining is not possible due to the unavailability of a gold standard, the lexicon-based approach provides comparable performance. We complemented the quantitative assessment of performance with the results of a qualitative error analysis. Lessons learned can be used to inform future research on emotion awareness in software engineering. In future work, we plan to extend the contributions of this study by performing a cross-study between datasets as a further performance evaluation technique.

## ACKNOWLEDGMENTS

This work is partially funded by the project 'EmoQuest - Investigating the Role of Emotions in Online Question & Answer Sites', funded by MIUR (Ministero dell'Università e della Ricerca) under the program "Scientific Independence of young Researchers" (SIR).